\documentclass[%
reprint,
showpacs,preprintnumbers,
amsmath,amssymb,
prb,
]{revtex4-1}

\usepackage{epstopdf}
\usepackage{graphicx}
\usepackage{dcolumn}
\usepackage{bm}
\usepackage{hyperref}
\usepackage{ulem}
\usepackage{color}

\graphicspath{ {./Figures//} }
\newcommand{\exc}{\langle N_{eh} \rangle}

\newcommand{\gf}{g^{(2)}}

\usepackage{xspace}

\begin{document}

\title{CdSe/CdS dot-in-rods nanocrystals fast blinking dynamics.}%

\author{M. Manceau$^{1,2}$, S. Vezzoli$^{3}$, Q. Glorieux$^1$,
E. Giacobino$^1$, L. Carbone$^5$, M. De Vittorio$^{4,5}$, J.-P. Hermier$^{6}$ and A. Bramati$^1$}

\affiliation{
 1 Laboratoire Kastler Brossel, UPMC-Sorbonne Universit\'es, CNRS, ENS-PSL Research University, Coll\`ege de France,
   4, place Jussieu Case 74, F-75005 Paris, France\\
 2 Universit\'e Paris 13, Sorbonne Paris Cit\'e, Laboratoire de Physique des Lasers, F-93430 
Villetaneuse, France\\  
 3 The Blackett Laboratory, Department of Physics, Imperial College London, London SW7 2AZ, United Kingdom\\
 4 Istituto Italiano di Tecnologia (IIT) Center for Bio-Molecular Nanotechnologies Via Barsanti sn, 73010 Arnesano 
(Lecce), Italy\\
 5 CNR NANOTEC- Institute of Nanotechnology c/o Campus Ecotekne, University of Salento, Via Monteroni - 73100 Lecce, 
Italy\\
 6 Groupe d’\'Etude de la Mati\`ere Condens\'ee, Universit\'e de Versailles-Saint-Quentin-en-Yvelines, CNRS UMR 8635, 
45, Avenue des Etats-Unis, F-78035 Versailles, France\\
}

\date{\today}

\begin{abstract}
The blinking dynamics of colloidal core-shell CdSe/CdS dot-in-rods is studied in detail at the single particle level. 
Analyzing the autocorrelation function of the fluorescence intensity, we demonstrate that these nanoemitters are 
characterized by a short value of the mean duration of bright periods (ten to a few hundreds of microseconds). The 
comparison of the results obtained for samples with different geometries shows that not only the shell thickness is 
crucial but also the shape of the dot-in-rods. Increasing the shell aspect ratio results in shorter bright periods 
suggesting that surface traps impact the stability of the fluorescence intensity.
\end{abstract}


\maketitle

\section{Introduction}\label{sec:Intro}

The emission intermittency, commonly called blinking, is characteristics of single nanocrystals emission. Since the 
first measurements on single CdSe nanocrystals\cite{nirmal1996} reporting a switching between an ON state emitting 
photons and an OFF state completely dark, this phenomenon has been the focus of intense studies because it deeply 
undermines the possible applications foreseen for these emitters, ranging from bio-imaging, light harvesting to 
nanophotonics\cite{geng2016} and quantum optics\cite{lounis2005,shcherbina2014}. A complete physical picture of the 
phenomenon has not been reached yet owing to the complexity of the processes at stake. 

Fluorescence blinking has been observed for various types of single nanoscale emitters\cite{frantsuzov2013}, including 
molecular dyes, fluorescent proteins, small nanodiamonds\cite{bradac2010} and colloidal nanocrystals. For complex 
single 
emitters such as CdSe nanocrystals, a broad distribution of blinking rates is observed, resulting in periods with a 
large 
fluorescence (on-state) and low fluorescence (off-state corresponding to the noise level) spanning from microseconds to 
hundreds of seconds\cite{kuno2000,Brokmann2003}.  Since the first report of fluorescence blinking in small spherical 
nanocrystals\cite{nirmal1996}, this behavior has been observed for many morphologies including elongated nanorods and 
nanowires. In the case of nanocrystals,  power law distributions\cite{kuno2000} with exponent smaller than $1$ were 
reported for cumulative durations of ON and OFF events.  These so called Levy distributions, have singular 
statistical properties: no mean value or standard deviation can be defined. Moreover long blinking periods are very 
probable as the decay of the distribution is slow. Also, more puzzling phenomena are associated with these 
distributions, such as statistical aging and non ergodicity\cite{Brokmann2003}. 

The last years have seen considerable progresses in reducing the effects of blinking thanks to new chemical synthesis 
methods\cite{mahler2008,chen2008} enabling the growth of CdS thick shells around the CdSe emitting core. It results in a better confinement of the 
charges inside the nanocrystal and the strong modification of the flickering dynamics. Long low-emitting periods are no more observed. Their duration 
does not exceed 100 ms. In addition, the emission does not turn completely off  and ``grey'' states have been identified \cite{spinicelli2009}. In 
that sense some articles mention non-blinking nanocrystals, even if super-Poissonian intensity fluctuations 
remain\cite{spinicelli2009,malko2011,galland2011,galland2012} indicating flickering between at least two emission levels. 

Beyond the growth of thicker shells, the overall nanocrystals shape (core and shell) can now be controlled. For example, 
dot-in-rods\cite{talapin2003,talapin2007,carbone2007} (DRs) consisting of a spherical core embedded in a cylindrical shell have been fabricated as 
well as nanoplatelets\cite{ithurria2011}. The modification of the shape opens new opportunities for the understanding and engineering of nanocrystals 
optical and spectroscopic properties\cite{louyer2011,tessier2012,biadala2014tuning,vezzoli2015}, in particular in view of pure single photon emission.

In this paper, we show that geometry plays an important role in the blinking dynamics of DRs and we propose a method of analysis for the blinking. We 
first observe that these DRs are characterized by a fast blinking dynamics. From a methodological point of view, our results indicate that it is 
usually poorly resolved when binning the signal, even with bin times as short as hundreds of microseconds. Due to the fast blinking dynamics, the 
common blinking analysis\cite{kuno2000,Brokmann2003} based on binning the photon detection events cannot yield any trustworthy information. In order 
to overcome this problem we show that a better approach consists in measuring the intensity autocorrelation function. In particular, it provides the 
average duration of bright periods. We then compare DRs samples with different geometries and demonstrate that, in addition to the shell thickness, 
the aspect ratio of the DR influences crucially the bright period duration. 

\section{Fast blinking dynamics}\label{sec:Fast}

For our study we used high quality CdSe/CdS core-shell DRs synthesized using the seeded growth approach proposed in 
reference\cite{carbone2007,pisanello2013}. In Tab.\ref{tab:Geometry} we give the core diameters, shell thicknesses and shell lengths of the various 
samples under study. The DRs are characterized by a minimum number of CdS monolayers on top of the core which is also given in 
Tab.\ref{tab:Geometry}. 
In the following DR1 corresponds to the thin shell dot-in-rods sample, while DR2, DR3 and DR4 are thick shell samples. For each sample a dilute 
toluene solution is drop-cast on a microscope glass coverslip to produce a low density of single DRs (typically $2$  DRs per $5$~$\mu$m$^2$ area). A 
single DR can be  chosen and excited using a picosecond-pulsed laser diode with a small excitation spot of $1$~$\mu$m$^2$. The picosecond-pulsed 
laser 
operates at a wavelength of $405$~nm and excites the highly absorptive shell \cite{carbone2007}, with a repetition rate of $2.5$~MHz. The 
photoluminescence (PL) is collected using a confocal microscope with a high numerical aperture objective ($100 \times$, N.A.=$1.4$). A high pass 
filter (cutoff $570$~nm) removes the remaining excitation light while leaving the DRs PL which is centered around $600-650$ nm for the various 
samples. The DRs PL is then spatially filtered through a pinhole and subsequently recorded using two single-photon avalanche photodiodes in a 
Hanbury-Brown and Twiss configuration.  The signals from the photodiodes were recorded by a Time-Correlated Single Photon Counting  data acquisition 
card enabling for each DR the recording of the PL autocorrelation function. Prior to any measurements we therefore check if the chosen particle is 
single or not with an antibunching measurement. 

\begin{figure}
  \centering
\includegraphics[scale=0.92]{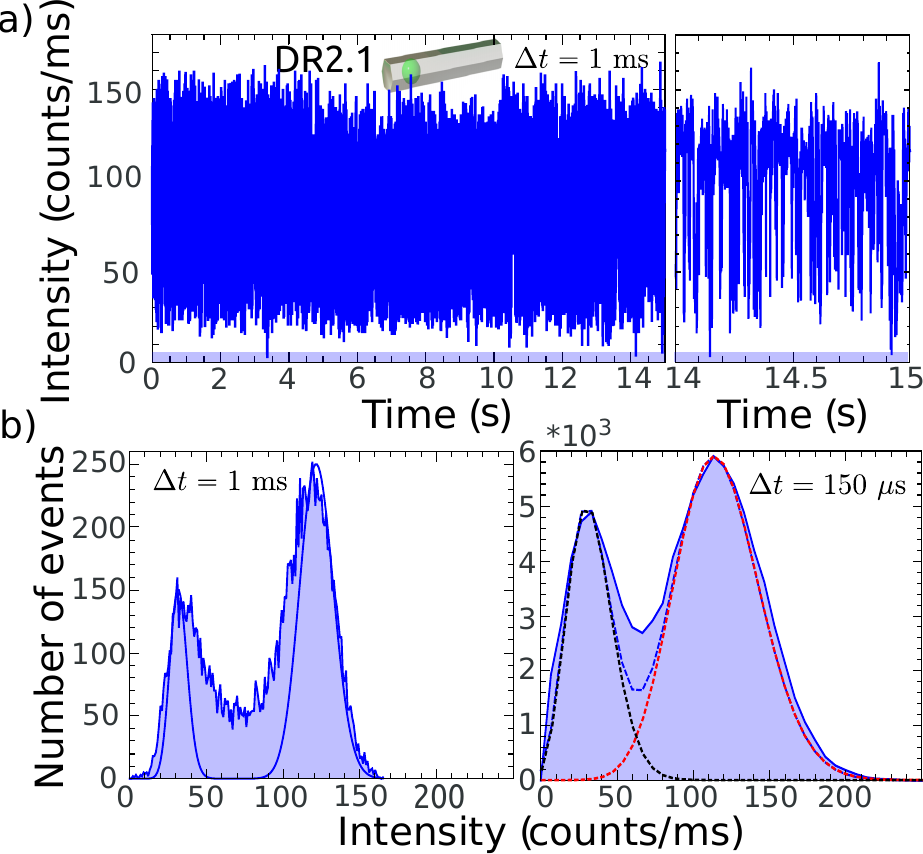}
  \caption{a) Left: typical PL timetrace of DR2.1 from sample DR2. Right: close view on the last second 
of the registered timetrace showing the flickering between two states. Mean excitation: $\exc=0.5$. Bin time $\Delta 
t=1$~ms. The noise level is given by the faint blue area.  b) Histogram of emission corresponding to the 
PL timetrace in a) for two time bins $\Delta t$. Each histogram is renormalized in counts/ms. Left: 
$\Delta t=1$~ms, , right: $\Delta t=150$~$\mu$s. A fit with the sum of two Poisson distributions (dashed blue line) is 
given in each case.} \label{fig:BinEffect} 
\end{figure}

We present the PL timetrace of a typical dot-in-rod of the thick shell sample DR2 in Fig.\ref{fig:BinEffect}a for an excitation below saturation. In 
the following  In the following we call DR2.1 this particular DR from sample DR2. The average number of electron-hole pairs $\exc$ inside the 
structure was measured to be $\exc=0.5$ by a saturation measurement\cite{Vezzoli2013}. The bin time is $\Delta t=1$~ms. The corresponding histogram 
of 
emission is shown in Fig.\ref{fig:BinEffect}b left. The histogram of emission for a PL timetrace computed with a $\Delta t=150$~$\mu$s bin time is 
also shown in Fig.\ref{fig:BinEffect}b right. This histogram reveals the presence of two emission peaks as expected for these 
emitters\cite{manceau2014} for an excitation under saturation. Indeed, DRs with a thick shell such as DR2.1 have a reduced blinking between a bright 
and a grey state, as demonstrated using time resolved decay measurements in one of our previous work \cite{manceau2014}, corresponding 
respectively  to the radiative recombination of an exciton and a negative trion\cite{spinicelli2009,galland2011} with a lower emission rate. The low 
emission state is well above the noise level here ($3$~count/ms shown in faint blue on Fig.\ref{fig:BinEffect}a). The intensity of emission of the 
bright and grey states are $I_{X}=116$~Counts/ms and $I_{X^-}=33$~Counts/ms respectively.  The histograms of emission can be fitted with the sum of 
two Poisson distributions (dashed blue line). In principle each emission state should correspond to a single Poisson distribution once the signal is 
properly binned. However Fig.\ref{fig:BinEffect}b clearly shows the limitations of such a fitting procedure.

One can see in Fig.\ref{fig:BinEffect}a that the emission is characterized by a fast switching between the two states. Indeed the $\Delta t=1$~ms bin 
time in Fig.\ref{fig:BinEffect}b left can only poorly resolve the emission dynamics. A broad range of intermediate emission intensities is visible in 
between the two emission peaks because of the time averaging imposed by the binning of the data. The binning of $\Delta t=150$~$\mu$s is far more 
accurate in resolving the emission dynamics as a better although, not perfect agreement is found with the two state emission fit. When using a 
technique of analysis of the signal relying on binning the photon detection events together, it is important to optimize the value of the time bin. 
The question of finding an ``optimized'' bin time is intrinsically linked to the timescales at which the blinking process occurs.  In the 
nanocrystals 
literature, most of the publications present data with bin times of $10$~ms or more. Large time bins are not suitable for the DRs under 
investigation. 
However, it is also important to state that the bin time cannot be set to extremely short values. Indeed, one is also limited by the photon 
collection 
rate, here a $\simeq 100$~counts/ms for the chosen excitation. The shorter the bin time the fewer detection events per time bin and the broader the 
corresponding Poisson distribution. Short time bins lead therefore to overlapping distributions as is visible in Fig.\ref{fig:BinEffect}b right for 
$\Delta t=150$~$\mu$s where the two Poisson distribution clearly overlap. Ultimately, a bin time of the order of the emitter lifetime will lead to on 
average less than one photon per time bin, with grey and bright states becoming completely indistinguishable. Indistinguishable grey and bright 
states 
due to overlapping distributions are a problem when a threshold needs to be set to distinguish the two states as will be seen in the next section. 

\section{Characterizing the blinking dynamics}\label{sec:FastBlinking}

\begin{figure}[h!]
  \centering
\includegraphics[scale=0.92]{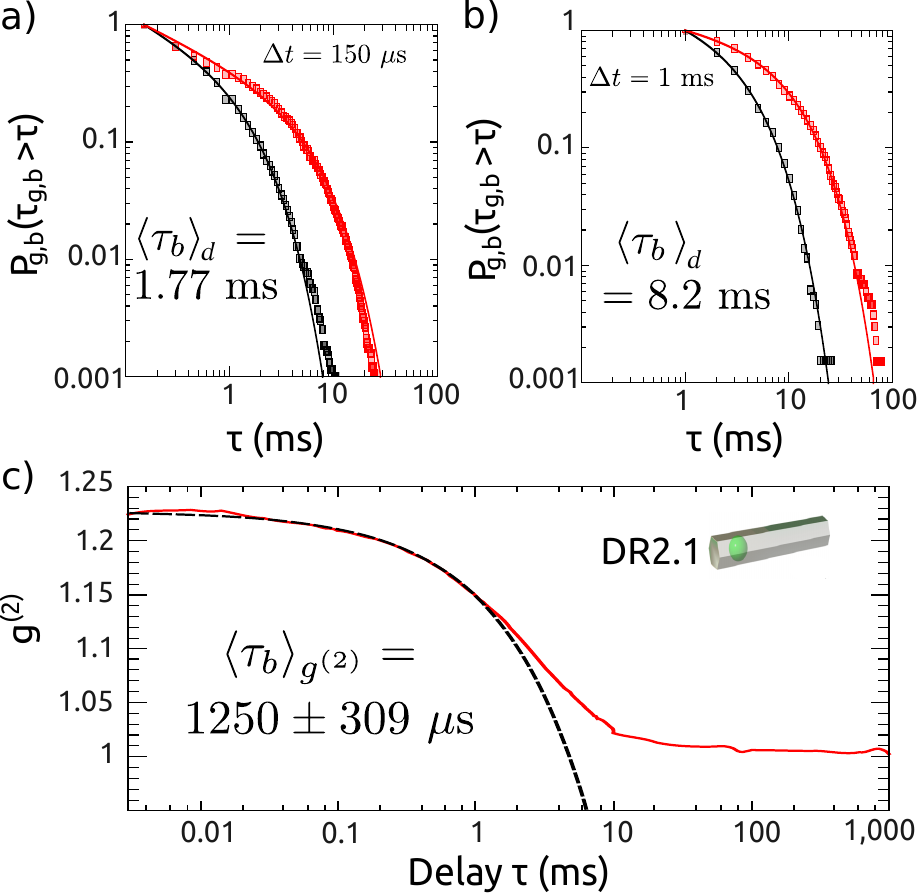}
  \caption{ 
a) Cumulative distributions of the bright and grey states event durations for DR2.1 corresponding to the histogram 
presented on 
Fig.\ref{fig:BinEffect}b right. Red: bright state, black: grey state. Bin time $\Delta t=150$~$\mu$s, thresholds: 
$I_b=I_g=60$~counts/ms. Full lines: power law distribution with exponential cutoff fit (red: $\mu=0.42$ and 
$\tau_c=6.1$~ms, black: $\mu=0.46$ and $\tau_c=1.5$~ms), see Eq.\ref{eq.PowerLawExpCutoffDistrib}.  b)  Cumulative 
distributions of the bright and grey states event durations corresponding to the histogram presented on 
Fig.\ref{fig:BinEffect}b left. Red: bright state, black: grey state. Bin time $\Delta t=1$~$m$s, thresholds: 
$I_b=I_g=60$~counts/ms. Full lines: power law distribution with exponential cutoff fit (red: $\mu=0.16$ and 
$\tau_c=10.5$~ms, black: $\mu=0.24$ and $\tau_c=3.8$~ms), see Eq.\ref{eq.PowerLawExpCutoffDistrib}. c) $\gf$ function 
for the detection events presented in Fig.\ref{fig:BinEffect}. Black dashed line: fit using Eq.\ref{eq.g2PowerLaw}. 
This gives $\langle \tau_b \rangle_{\gf}=1250\pm309$~$\mu$s and $\mu=0.34\pm0.036$.}
  \label{fig:ON-OFFAnalysis}
\end{figure}

We now present a more quantitative analysis of the reduced blinking dynamics of thick shell DRs. In Fig.\ref{fig:ON-OFFAnalysis}a and 
Fig.\ref{fig:ON-OFFAnalysis}b we present the cumulative distributions of the bright ($\mathcal{P}_{b}(\tau_{b}\geq\tau)$)  and grey 
($\mathcal{P}_{g}(\tau_{g}\geq\tau)$) states event durations from the DR2.1 timetrace presented on Fig.\ref{fig:BinEffect}a. The cumulative 
distribution $\mathcal{P}_{b,g}(\tau_{b,g}\geq\tau)$ as a function of $\tau$ gives the probability that the bright (grey) period $\tau_b$ ($\tau_g$) 
is larger than $\tau$. Bin times of $\Delta t=150$~$\mu$s and $\Delta t=1$~$m$s are used respectively. The thresholds $I_b$ and $I_g$ for the bright 
and grey states events are fixed in between the two states at $I_b=I_g=60$~counts/ms. For the case of $\Delta t=150$~$\mu$s, this threshold value 
corresponds roughly to a distance of $5$ standard deviation to the mean value of each emission state. This way a minimum of overlap between the 
states 
is ensured as can be seen from the fit in Fig.\ref{fig:BinEffect}b right and no data is discarded from the analysis in this case. Time bins with 
intensities above $I_b$ are considered as part of the bright state and time bins with intensities below $I_g$ are considered as part of the grey 
state. The cumulative distributions shows that long periods are strongly inhibited for both type of events, grey or bright as expected from the 
timetrace in Fig.\ref{fig:BinEffect}a. Fits of the cumulative distributions are presented in Fig.\ref{fig:ON-OFFAnalysis}a as full lines 
corresponding 
to power laws with exponential cutoff:

\begin{equation}\label{eq.PowerLawExpCutoffDistrib}
\mathcal{P}(\tau_{b,g}\geq\tau)\propto \frac{1}{\tau^\mu}e^{-\tau/\tau_c},
\end{equation}

\noindent with $\mu$ the power law exponent and $\tau_c$ the exponential cutoff time. For $\Delta t=150$~$\mu$s, we obtain power law exponents of 
$0.46$ and $0.42$ for the grey and bright states respectively with exponential cutoffs of $1.5$ and $6.1$~ms for the cumulative distributions. 
Here we report values of the power law exponent $\mu$ corresponding to cumulative distributions. Hence the values close to $0.5$ 
\cite{Brokmann2003} corresponds to $1+\mu\simeq1.5$ often reported in the literature for the non-cumulative distributions.

However it should be noted that in the case of two states with close emission rates, the analysis of event durations is not totally reliable, indeed 
it strongly depends on the thresholds chosen and also on the bin time $\Delta t$ of the intensity timetrace as demonstrated in reference 
\cite{crouch2010,amecke2014}. In Fig.\ref{fig:ON-OFFAnalysis}b we present the same analysis with a larger bin time than in 
Fig.\ref{fig:ON-OFFAnalysis}a, $\Delta t=1$~ms,  and the same thresholds. The curves have the same shapes, the fits yield power law exponents of 
$0.24$ and $0.16$ for the grey and bright states respectively with exponential cutoffs of $3.8$ and $10.5$~ms. Changing the bin time has considerably 
modified the distributions and the fitting values.  We already know that the bin time of $\Delta t=1$~ms is less relevant that the bin time of 
$\Delta 
t=150$~$\mu$s for the studied timetrace as explained in the previous section. Many events in the PL timetrace have durations between $150$~$\mu$s and 
$1$~ms as attested by the slope of the cumulative distribution on these timescales in Fig.\ref{fig:ON-OFFAnalysis}a. Indeed, approximately $63\%$ of 
the events considered as bright with the  bin time of $\Delta t=150$~$\mu$s have durations smaller than $\tau=1$~ms as 
$\mathcal{P}_{b}(\tau_{b}\geq1$~ms$)=37\%$ in Fig.\ref{fig:ON-OFFAnalysis}a. Hence all these events cannot be resolved with a bin time of $\Delta 
t=1$~ms. Furthermore, bright events with duration shorter than a $150$~$\mu$s also exist, but cannot be grasped by the $\Delta t=150$~$\mu$s binning. 
This explains the imperfect fitting by the two Poisson distributions in Fig.\ref{fig:BinEffect}b right. 

In the following we will investigate the blinking between the bright and grey states using a different approach: the intensity correlation 
function\cite{bernard1993,fleury2000,messin2001,verberk2002,pelton2004,pelton2007,canneson2014,houel2015}.  The autocorrelation method is less 
straightforward than the distribution of event durations, but it does not suffer from any a priori assumptions due to the time bin and it yields 
information at short timescales not reachable when binning the signal. 

Fig.\ref{fig:ON-OFFAnalysis}c presents the $\gf$ function corresponding to the PL timetrace in Fig.\ref{fig:BinEffect}. 
The temporal intensity correlation $\gf$ of a light field  is defined as follows: 

\begin{equation}\label{eq.g2-Classical-1} 
g^{(2)}(t_0,t_0+\tau)=\frac{\langle I(t_0)I(t_0+\tau)\rangle}{\langle I(t_0)\rangle^2}, 
\end{equation}

\noindent with $t_0$, $t_0+\tau$ times, $\langle\rangle$ denotes a statistical (ensemble) average.  For the DRs presented in this paper switching 
between two emission states characterized by  power law distributions with exponential cutoffs of the event  durations 
(eq.\ref{eq.PowerLawExpCutoffDistrib}), an expression for the $\gf$ function is derived in references \cite{verberk2002,verberk2003}. Such a model 
can 
be applied to investigate the flickering of CdSe/CdS colloidal emitters since the duration of the low emitting periods does not exceed tens of 
milliseconds \cite{mahler2008}, meaning that a cutoff is always observed as can be seen in Fig.\ref{fig:ON-OFFAnalysis}a. On timescales smaller than 
the exponential cutoffs of the order of a couple of milliseconds the $\gf$ function can be expressed as:

\begin{equation}\label{eq.g2PowerLaw}
\gf(\tau)=B(1-A\tau^{1-\mu}).
\end{equation}

\noindent Here, $\mu$ is the largest power law exponent among the two states power law distributions. $B$ is the bunching value, \textit{i.e.} the 
value of $g^{(2)}$ in eq.\ref{eq.g2PowerLaw} at short timescales. Here the model does not take into account antibunching and single photon 
emission\cite{manceau2014} at short timescales  of the order of the emitter lifetime (nanoseconds to hundreds of nanoseconds). Therefore, as 
$\lim\limits_{\tau \to 0} \tau^{1-\mu}=0$, $\gf(0)$ is equal to the bunching value $B$ in eq.\ref{eq.g2PowerLaw}. For molecules and quantum dots with 
simple exponential blinking laws\cite{bernard1993,pistol1999,fleury2000,sychugov2005}, the bunching value corresponds to the ratio between the 
average 
OFF and ON periods. More generally, the fact that $g^{(2)}$ is larger than one on some timescales means that bunches of photons with various emission 
rate coexist. This leads to intensity fluctuations larger than Poisson fluctuation corresponding to a single rate of emission and consequently 
$g^{(2)}>1$. Furthermore, as shown in reference \cite{verberk2003} the parameter $A$ is equal to:

\begin{equation}\label{eq.A}
A=\frac{1}{ \langle \tau_{b} \rangle }\frac{\tau_{min}^{\mu}}{\Gamma(2-\mu)},
\end{equation}

\noindent with $ \langle \tau_{b} \rangle $ the average duration of the bright state events, $\tau_{min}$ the minimum duration of a event and 
$\Gamma$ 
the usual gamma function. For our experiment, the minimum duration of an emission event $\tau_{min}$ is equal to the measurement time 
resolution\cite{pelton2004}, $400$~ns, corresponding to the delay between two excitation pulses.  The factor $A$ is inversely proportional to the 
average duration of the bright state. Hence $ \langle \tau_{b} \rangle $, together with $\mu$, defines on which timescales the $\gf$  falls towards 
unity as the blinking correlations are lost due to the cutoffs at long time scales.

In the following, we use the notation $\langle \tau_b \rangle_{d}$ and $\langle \tau_b \rangle_{\gf}$ for the average bright period duration obtained 
from the blinking distribution and the correlation function respectively. $\langle \tau_b \rangle_{d}$ is the mean value obtained from the 
experimental probability distributions, not the cumulative distributions presented in this article.  The probability distribution of bright events 
for 
DR2.1 corresponding to the cumulative distribution presented in Fig.\ref{fig:ON-OFFAnalysis}a yields $\langle \tau_b \rangle_{d}=1.77$~ms with 
$I_b=60$~counts/ms and $\Delta t=150$~$\mu$s. In order to determine $\langle \tau_b \rangle_{\gf}$ we fitted the $\gf$ curve in 
Fig.\ref{fig:ON-OFFAnalysis}c by Eq.\ref{eq.g2PowerLaw} and \ref{eq.A}(see fitting method in Appendix \ref{App1:fitting}for more details). We find 
$\langle \tau_b \rangle_{\gf}=1250\pm309$~$\mu$s, the $309$~$\mu$s uncertainty being the fitting error (see Appendix \ref{App1:fitting}). The value 
is 
smaller than the one found with the blinking distribution. This can be explained by the fact that the blinking distribution does not take into 
account 
the fast events (faster than the bin time of $\Delta t=150$~$\mu$s). Hence and as already stated in reference \cite{kuno2000}, the average times 
deduced from the blinking distributions depend on the bin time chosen because of the scale invariance of the power law. It is interesting to state 
that the result of a mixing between the two states due to a poor bin or threshold choice is a longer average time, even though grey periods are on 
average shorter than bright periods (the grey periods distribution is under the bright periods distribution in Fig.\ref{fig:ON-OFFAnalysis}a). For 
example  $\langle \tau_b\rangle_d=1.77$~ms was obtained with  the threshold $I_b=60$~counts/ms and $\Delta t=150$~$\mu$s. For the case of a bin time 
$\Delta t=1$~$m$s in Fig.\ref{fig:ON-OFFAnalysis}b, the distribution of bright period gives a very poor estimation of the average bright period: 
$\langle \tau_b \rangle_d=8.2$~ms. This can be easily understood as the mixing between the two states tends to create long periods with the same 
intensities. Ultimately, a very large bin would give an average event duration of the order of the measurement time.

For the specific nanocrystal DR2.1 of Fig.\ref{fig:BinEffect} and Fig.\ref{fig:ON-OFFAnalysis}, an appropriate choice of bin time can  thus fairly 
well resolve the blinking dynamics as the average switching time between the two states is large enough. A rough estimation of the blinking dynamics 
can be made using the distribution of bright and grey blinking periods although the results is still biased from the choice of a threshold. In the 
case of the DR2.1, one should also note that $\langle \tau_b \rangle_{\gf}=1250 \mu s$ is close to the time scale for which the fit provided by eq. 
\ref{eq.A} is valid. This reduces the accuracy of the$\langle \tau_b \rangle_{\gf}$ value.

In the next section we show some examples for which the use of the distributions of bright and grey blinking periods is irrelevant. The 
autocorrelation function is then necessary. Various DR samples are studied and we demonstrate that the nanocrystals geometry has an impact on the 
blinking dynamics.

\section{Blinking dynamics and nanocrystal geometry}\label{sec:Blinking&Geometry}

\begin{figure}
  \centering
\includegraphics[scale=0.92]{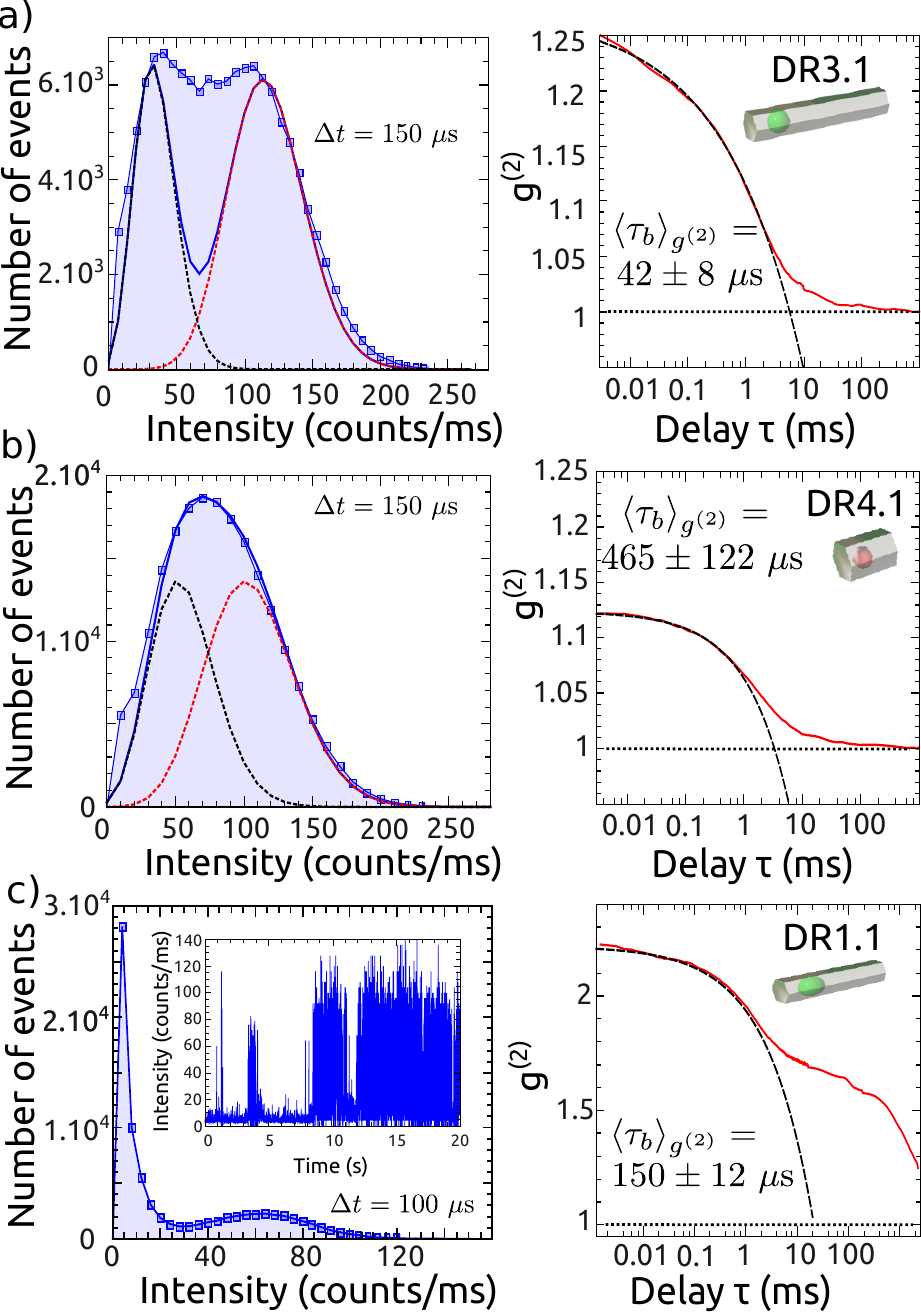}
  \caption{Histograms and $\gf$ function for DR3.1, DR4.1 and DR1.1 from samples DR3, DR4 and DR1 respectively, excited 
at $\exc\leq0.5$. Right panels,
red curve: $\gf$ function, black dashed curve: fit with eq.\ref{eq.g2PowerLaw}. a) DR3.1, bin time $\Delta 
t=150$~$\mu$s,  $\langle \tau_b 
\rangle_{\gf}=42\pm8$~$\mu$s, and $\mu=0.69\pm0.025$ . b) DR4.1, bin time $\Delta t=150$~$\mu$s,  $\langle \tau_b 
\rangle_{\gf}=465\pm122$~$\mu$s, and $\mu=0.49\pm0.034$. c) DR1.1, bin time $\Delta t=100$~$\mu$s,  $\langle \tau_b 
\rangle_{\gf}=150\pm12$~$\mu$s and $\mu=0.52\pm0.01$.}
  \label{fig:BlinkingThickShells} 
\end{figure}

Fig.\ref{fig:BlinkingThickShells} presents the histograms of emission and $\gf$ functions for DR3.1, DR4.1, and DR1.1, three representatives DRs of 
respectively samples DR3, DR4, and DR1, for approximately the same mean excitation $\exc=0.5$ as in the former case of DR2.1. One can see that the 
blinking dynamics of DR3.1 should be characterized by short bright and grey periods. Indeed the histogram of emission in 
Fig.\ref{fig:BlinkingThickShells}a left has a broad intermediate intensity range in between the two emission peaks due to the bin averaging even 
though a short bin time of $\Delta t=150$~$\mu$s is used. The analysis in terms of cumulative distributions of blinking periods is inappropriate in 
this case. The fast blinking dynamics can nevertheless be quantitatively estimated by the fit of the $\gf$ function in 
Fig.\ref{fig:BlinkingThickShells}a right. It yields an average bright period duration twenty times shorter than for DR2.1 (Fig.\ref{fig:BinEffect} 
and 
Fig.\ref{fig:ON-OFFAnalysis}) with $\langle \tau_b \rangle_{\gf}=42\pm8$~$\mu$s. 

The $\gf$ curve for DR4.1 appears to be almost flat for $\tau<500$~$\mu$s in Fig.\ref{fig:BlinkingThickShells}b right as for DR2.1 in 
Fig.\ref{fig:ON-OFFAnalysis}c. The average bright periods duration is long compared to DR3.1, $\langle \tau_b \rangle_{\gf}=465\pm122$~$\mu$s is 
found 
through the fit of the intensity correlation function. The switching dynamics is therefore on average slower than for DR3.1. The intensity histogram 
with a $\Delta t=150$~$\mu$s bin time is well represented by the sum of two Poisson distributions in Fig.\ref{fig:BlinkingThickShells}b left. It is 
to 
be noted that the two states of emission overlap on a larger intensity range than for the previous examples. DR4.1 has a very thick shell and 
therefore a larger trion quantum yield\cite{galland2012,malko2011}. In this case the estimation of the blinking statistics with the cumulative 
distributions is  inappropriate\cite{crouch2010} as the two states distributions largely overlap and the grey and bright photons are mixed when 
binning the signal. 

We also present the results for thinner shell dot-in-rods DR1.1 from sample DR1. In contrast to the previous thick shells samples, this sample 
switches between a bright and a dark (noise level) state rather than grey state. Also it sometimes shows long dark periods of some seconds, while the 
bright periods are as previously always limited to a maximum of a few hundreds of microseconds. This is clearly visible in 
Fig.\ref{fig:BlinkingThickShells}c left that shows the PL (inset) and the corresponding intensity histogram of such a DR with a bin time of $\Delta 
t=100$~$\mu$s. The corresponding $\gf$ has characteristics similar to the other DRs correlation functions at short timescales ($<10$~ms). This is due 
to the part of the PL timetrace measured after $10$~s of PL recording for which a fast switching dynamics is observed in 
Fig.\ref{fig:BlinkingThickShells}c left inset. The intensity correlation function nevertheless displays an additional bunching tail at larger 
timescales ($\tau>10$~ms) due to the long blinking events characteristics of thin shell nanocrystals. These long blinking events are visible 
in the first $10$ seconds of the PL recording. The additional decreasing tail of the intensity correlation function for $\tau>500$~ms is due to the 
finite acquisition time. The fit of the $\gf$ curve in Fig.\ref{fig:BlinkingThickShells}c right gives $\langle \tau_b \rangle_{\gf}=150\pm12$~$\mu$s. 
Let us also note that for these DRs showing fast blinking, the value of  $\langle \tau_b \rangle_{\gf}$ is also very accurate since the agreement 
between the experimental results and the fit is good for time scales much higher than $\langle \tau_b \rangle_{\gf}$.

Finally, we present an analysis of the PL of more than $40$ DRs for each sample presented previously. The excitation 
was kept below saturation, in between $\exc=0.1$ and  $\exc=0.5$ for each DR to avoid the excitation of higher 
order states and to stay in the two states blinking regime\cite{manceau2014,galland2011,galland2012}. The intensity 
correlation function was computed for each DR and fitted with Eq.\ref{eq.g2PowerLaw} following the procedure described 
in App.\ref{App1:fitting}. The sample mean values of the average bright period duration is reported for each sample in 
Tab.\ref{tab:Geometry}. The samples mean values are within a range of $180$ to $400$~$\mu$s. Our DRs are therefore 
characterized by a fast switching dynamics that can be only poorly resolved when binning the signal on hundreds 
microseconds to milliseconds as previously stated. A large dispersion of values exists within each sample, with some 
DRs having average bright periods of a few tenth of microseconds but also up to $1$~ms as presented in Fig.2. This 
might be due to a slight dispersion in sizes within a given sample and different electrostatic environment between 
single dots. The histograms of the various values of bright periods found for each sample are given in 
App.\ref{App2:Distributions}. It is apparent on these histograms that samples DR1 and DR3 have more single DRs with 
very short average bright periods (shorter than $100$~$\mu$s) than samples DR2 and DR4.

We can also notice that even though samples DR1, DR2 and DR3 have the same core, they have different average bright time durations. As expected, the 
comparison between samples DR1 and DR2 shows that an increase of the shell thickness results in a decrease of the switching dynamics from the bright 
state to the grey one. However, the DR shell aspect ratio, \textit{i.e.} the length of the shell over its thickness, is also crucial. Even if samples 
DR2 and DR3 exhibit the same core diameter and shell thickness, Table \ref{tab:Geometry} indicates that the increase of the DR length deteriorates 
the bright state stability. Surface  trap states on the shell may be responsible of this observation. A larger shell aspect ratio leads to a faster 
blinking dynamics. Sample DR3, with a shell aspect ratio of $8.3$, has an average bright period duration $1.6$ times shorter than sample DR2 with a 
shell aspect ratio of $3.1$. The results concerning sample DR4 confirm the previous observations pointing towards surface trap states as responsible 
for the instability of the bright state rather than traps inside the nanocrystals volume. Indeed, the synthesis of a very thick shell enables to 
significantly increase $\langle\tau_b\rangle$ beyond the value obtained for sample DR2 that exhibits approximately the same aspect ratio as sample 
DR4. 


\begin{table*}[t]
\centering
\begin{tabular}{| c | c | c | c | c | c | c | c | c | c |}
  \hline
  & \shortstack{Core diameter \\ (nm)}  &  \shortstack{Thickness \\ (nm)} & \shortstack{\hphantom{DR1} \\ Length \\ 
(nm)}  & 
  \shortstack{Aspect ratio \\ \hphantom{DR1}} &  \shortstack{ CdS monolayers  \\ on top of CdSe core} & 
\shortstack{$\langle\tau_b\rangle_{\gf}$ ($\sigma$) \\ \hphantom{DR1}}  \\
  \hline
  \hline
  DR1& $3.3$ &$4$ & $22$ & $5.5$ & $1$ & $190$~$\mu$s ($189$) \\
  \hline
  DR2& $3.3$ &$7$ & $22$ &  $3.1$ & $4$ & $299$~$\mu$s ($264$)\\
  \hline
  DR3& $3.3$ &$7$ & $58$ & $8.3$ & $4$ & $186$~$\mu$s ($175$) \\
  \hline
  DR4& $4.6$ &$11$ & $29$ & $2.6$ & $8$ & $394$~$\mu$s ($299$) \\
  \hline
\end{tabular}
   \caption{Geometrical parameters of the investigated samples. The aspect ratio is the ratio between the length and the thickness of the shell. 
The last column gives the samples mean values and 
dispersions for the bright state average duration obtained from the autocorrelation function of more than $40$ DRs per 
sample. See Fig.\ref{fig:Histogramtb} in App.\ref{App2:Distributions} for the various samples 
$\langle\tau_b\rangle_{\gf}$ distributions.}
  \label{tab:Geometry}
\end{table*}

\section{Conclusion}\label{sec:Conclusion} In conclusion, an analysis of the fast blinking dynamics of single colloidal CdSe/CdS DRs was presented. 
This analysis relies on the autocorrelation function which is shown to be the most suitable tool to properly characterize the flickering between 
emission states. We showed that the average duration of the bright periods of such emitters is of the order of few hundreds of microseconds, it is so 
short that any method relying on binning the signal can not resolve appropriately the flickering dynamics while the autocorrelation function gives 
access to the full blinking dynamics. We also characterized the blinking of several DR samples. Our results demonstrate that not only the thickness 
of 
the shell but also its shape has to be considered. The decrease of the bright period duration with the aspect ratio of the shell suggests that traps 
at the surface of the DR are involved in the flickering of the emission.

\appendix

\section{fitting of the correlation function}\label{App1:fitting}

The normalized correlation functions $\gf$ were fitted by Eq.\ref{eq.g2PowerLaw}:

\begin{equation}
\gf(\tau)=B(1-A\tau^{1-\mu}),
\end{equation}

\noindent with $B$, $A$ and $\mu$ as free fitting parameters. $\langle \tau_{b}\rangle_{\gf}$ is subsequently 
determined for 
each nanocrystal following Eq.\ref{eq.A}:

\begin{equation}
A=\frac{1}{\langle \tau_{b}\rangle}\frac{\tau_{min}^{\mu}}{\Gamma(2-\mu)},
\end{equation}

\noindent with $\tau_{min}=400$~ns the measurement time resolution.  One unknown parameter is the long time exponential 
cutoff as defined in 
Eq.\ref{eq.PowerLawExpCutoffDistrib} that characterizes the distribution of bright and grey periods. 
Eq.\ref{eq.g2PowerLaw} is valid for 
correlation times shorter than the exponential cutoff\cite{verberk2002,verberk2003}. A limit $\tau_f$ to the fit has to 
be chosen for each 
DR. 

\begin{figure}[h]
  \centering
\includegraphics[scale=0.86]{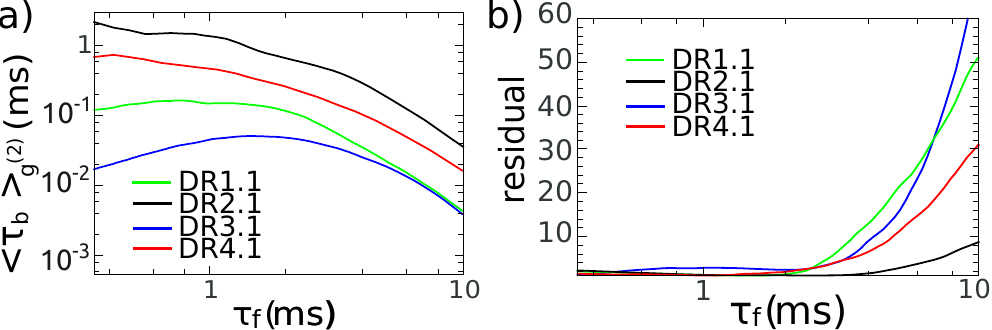}
  \caption{a) $\langle\tau_{b}\rangle_{\gf}$ against the fit limit $\tau_f$ for the $\gf$ curves presented in 
fig.\ref{fig:ON-OFFAnalysis} and  fig.\ref{fig:BlinkingThickShells}. Averaging the fitted values for $\tau_f<2$~ms for 
each DR 
gives: DR1.1 (green) $\langle \tau_b \rangle_{\gf}=150\pm12$~$\mu$s, DR2.1 (black) $\langle \tau_b 
\rangle_{\gf}=1250\pm252$~$\mu$s, DR3.1 (blue) $\langle \tau_b \rangle_{\gf}=42\pm8$~$\mu$s and DR4.1 (red) $\langle 
\tau_b 
\rangle_{\gf}=465\pm122$~$\mu$s  b) Fit normalized residual against fit threshold $\tau_f$. }
  \label{fig:Fitting} 
\end{figure}

Fig.\ref{fig:Fitting}b presents the normalized residuals for the  $\gf$ fits presented in fig.\ref{fig:ON-OFFAnalysis} 
and  fig.\ref{fig:BlinkingThickShells} against the fit limit $\tau_f$. The residual was normalized to its maximum value 
at $\tau_f=10$~ms. For  $\tau_f>10$~ms the residual increases and the fit becomes clearly inappropriate.

We can see that the residual stays almost constant up to roughly $\tau_f=2$~ms before increasing for larger cutoff as 
the model cannot fit the $\gf$ curves for larger delays. The corresponding $\langle \tau_{b}\rangle_{\gf}$ values 
against the fit limit $\tau_f$ are shown in Fig.\ref{fig:Fitting}a. The different curves shows that $\langle 
\tau_{b}\rangle_{\gf}$ is almost constant for $\tau_f<2$~ms before abruptly dropping while the fit residuals increase 
for $\tau_f>2$~ms. The $\langle \tau_{b}\rangle_{\gf}$ values reported for single nanocrystals in this article are the 
average values of the fit for $\tau_f<2$~ms, while the given fitting errors correspond to the standard deviation of the 
various fitted values over the range $\tau_f<2$~ms.

\section{Sample distributions of $\mathbf{\langle\tau_b\rangle_{\gf}}$}\label{App2:Distributions}

\begin{figure}
  \centering
\includegraphics[scale=0.90]{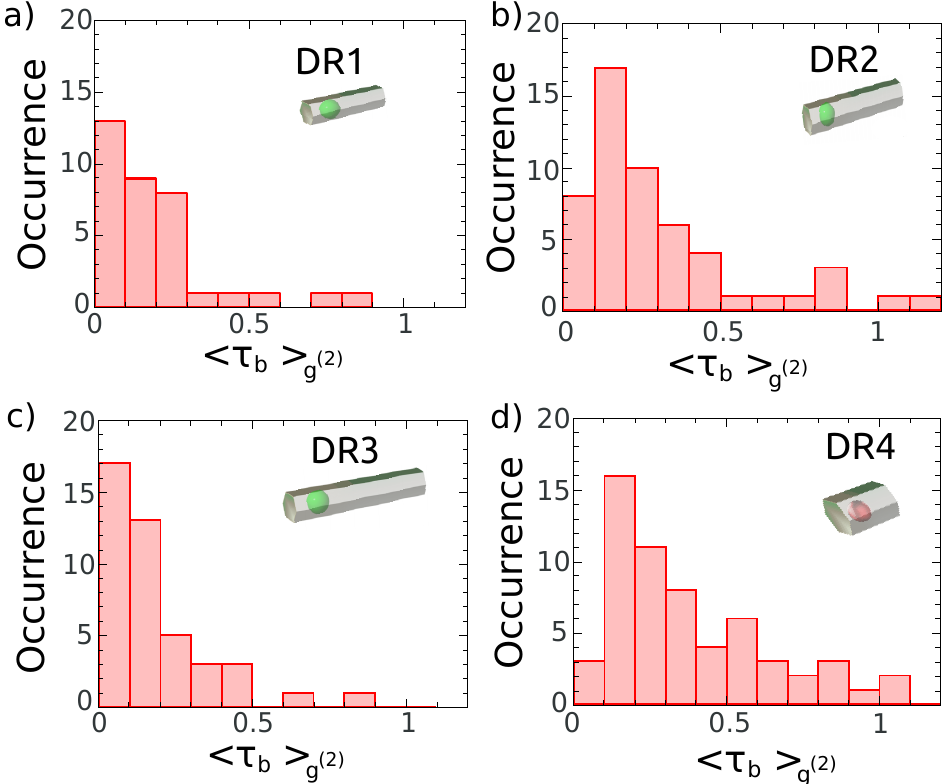}
  \caption{Histogram of the average bright period duration for each studied sample. The samples mean values of $\langle 
\tau_{b}\rangle_{\gf}$ are: a) DR1: $190$~$\mu$s ($\sigma=189$~$\mu$s),  b) DR2: $299$~$\mu$s ($\sigma=264$~$\mu$s), c) 
DR3: 
$186$~$\mu$s ($\sigma=175$~$\mu$s), d) DR4: $394$~$\mu$s ($\sigma=299$~$\mu$s).}
  \label{fig:Histogramtb} 
\end{figure}

Following the fitting procedure explained in Appendix \ref{App1:fitting} for each single nanocrystals, the sample 
distributions of average 
bright periods duration shown in Fig.\ref{fig:Histogramtb} were found by measuring the intensity correlation function of 
more than $40$ 
single nanocrystals per sample. The samples mean values and standard deviations reported in Tab.\ref{tab:Geometry} are 
taken from the 
samples distributions shown in Fig.\ref{fig:Histogramtb}.

\section{Excitation rate and $\mathbf{\tau_{min}}$}\label{App3:taumin}

The parameter $\tau_{min}$ in eq.\ref{eq.A} is the minimum duration of a bright or grey event. In reference 
\cite{verberk2003}, it is included in the model as the short timescale cutoff of the blinking duration distributions 
$\mathcal{P}(\tau_{b,g}\geq\tau)\propto \frac{1}{\tau^\mu}e^{-\tau/\tau_c}$. These distributions need a short time 
cutoff because they diverge for $\tau\to0$. Physically the blinking cannot be infinitely fast. We define the minimum 
duration of a bright or grey event $\tau_{min}$ in eq.\ref{eq.A} as the delay $\Delta$ between two excitation pulse  as 
proposed in reference \cite{pelton2004}, here $\tau_{min}=400$~ns. This is reasonable as long as the excitation 
repetition rate is low enough. Indeed, in this case the resulting delay between pulses is larger than the typical 
timescale of any physical mechanism that would prevent the switching between states. Therefore in this case $\tau_{min}$ 
is fixed by the excitation rather than by a physical mechanism.

\begin{figure}
  \centering
\includegraphics[scale=0.92]{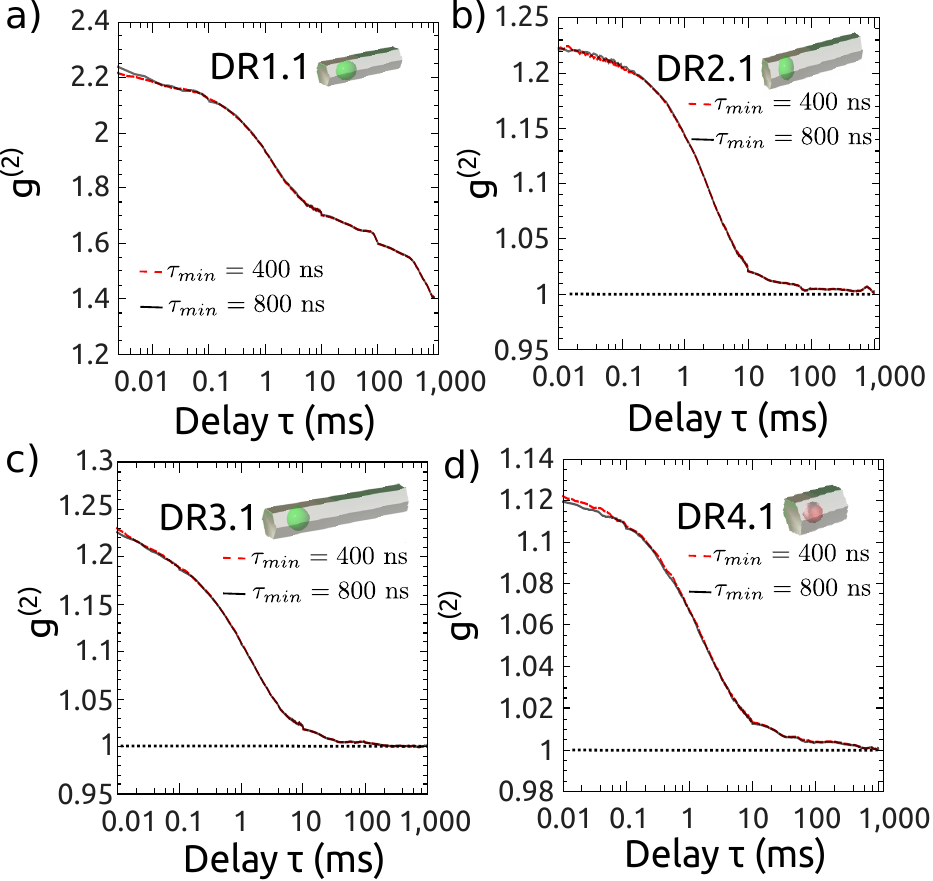}
  \caption{ $\gf$ function for DR1.1, DR2.1, DR3.1 and DR4.1 from samples DR1, DR2, DR3 and DR4 respectively. The red 
dashed curve is the $\gf$ function calculated from the raw data as already presented in fig.\ref{fig:ON-OFFAnalysis} 
and fig.\ref{fig:BlinkingThickShells} with $\tau_{min}=400$~ns. The solid black line is the $\gf$ function calculated 
from the raw data keeping only every second excitation such that $\tau_{min}=800$~ns. }
  \label{fig:tauMin} 
\end{figure}

The rate of excitation of a nanocrystal can heavily modify the blinking statistics, and thus the cumulative 
distribution of events $\mathcal{P}(\tau_{b,g}\geq\tau)$ and the corresponding $\gf$ function, as shown in reference 
\cite{schwartz2012}. Hence we do not consider a change of the rate of excitation experimentally because this might 
lead to compare blinking statistics for a given nanocrystal with different power law exponents $\mu$ for example. In 
this case one would not be able to sort out the effect of changing the minimum delay between single photons 
$\tau_{min}$ on the average blinking duration with any other potential changes in the blinking statistics. To test the 
effect of the repetition rate on the $\langle\tau_b\rangle_{\gf}$ values via the $\tau_{min}$ parameter in 
eq.\ref{eq.A}, we  decrease the repetition rate by removing detected photons in the post-measurement data analysis. By 
removing every second excitation from a measurement, we can artificially simulate 
a decrease by a factor of 2 of the repetition rate.

In fig.\ref{fig:tauMin} we present the $\gf$ function of the nanocrystals studied in fig.\ref{fig:ON-OFFAnalysis} 
and fig.\ref{fig:BlinkingThickShells} calculated on all the photons registered experimentally (red dashed curve), and on 
only every second excitation (solid black line). The $\gf$ functions are almost identical. This implies that the fits 
give $\langle\tau_b\rangle_{\gf,\tau_{min}=800~ns}\simeq\langle\tau_b\rangle_{\gf,\tau_{min}=400~ns}\times2^{\mu}$ 
according to eq.\ref{eq.A}. The average bright period duration increases while increasing the minimum blinking duration 
$\tau_{min}$ as fast blinking events are removed from the statistics. Table \ref{tab:tauMin} gives the average blinking 
duration found trough the fitting procedure for the four nanocrystals under study for both $\tau_{min}=400~ns$ and 
$\tau_{min}=800~ns$.

\begin{table}[t]
\centering
\begin{tabular}{| c | c | c |}
  \hline
   &  \shortstack{\hphantom{DR1} \\$\langle\tau_b\rangle_{\gf}$ \\ $\tau_{min}=400$~ns}  
   &  \shortstack{\hphantom{DR1} \\$\langle\tau_b\rangle_{\gf}$ \\ $\tau_{min}=800$~ns}   \\
  \hline
  \hline
  DR1.1& $150$~$\mu$s (12) &$213$~$\mu$s (23) \\
  \hline
  DR2.1& $1250$~$\mu$s (309) &$1595$~$\mu$s (478) \\
  \hline
  DR3.1& $42$~$\mu$s (8) &$86$~$\mu$s (15) \\
  \hline
  DR4.1& $465$~$\mu$s (122) &$698$~$\mu$s (211) \\
  \hline
\end{tabular}
   \caption{Average bright period duration $\langle\tau_b\rangle_{\gf}$ and its uncertainty found from a fit of the 
correlation function for DR1.1, DR2.1, DR3.1 and DR4.1 for $\tau_{min}=400$~ns and $\tau_{min}=800$~ns. 
$\tau_{min}=400$~ns corresponds to the raw data as presented in fig.\ref{fig:ON-OFFAnalysis} and 
fig.\ref{fig:BlinkingThickShells}, while $\tau_{min}=800$~ns corresponds the the same data for which every second 
excitation was removed.}
  \label{tab:tauMin}
\end{table}

\end{document}